\documentclass[fleqn]{article}
\usepackage{longtable}
\usepackage{graphicx}

\begin {document}
%\pagenumbering{arabic}
%\frenchspacing
%parindent 0.0 cm
%\parskip 0.5cm
%\begin{center}
\begin{flushleft}
{\LARGE
{\bf Discrepancies in Atomic Data and Suggestions for their Resolutions}
}\\

\vspace{1.5 cm}

{\bf {Kanti  M  ~Aggarwal}}\\ 

\vspace*{1.0cm}

Astrophysics Research Centre, School of Mathematics and Physics, Queen's University Belfast, \\Belfast BT7 1NN, Northern Ireland, UK\\ 
\vspace*{0.5 cm} 

e-mail: K.Aggarwal@qub.ac.uk \\

\vspace*{0.20cm}

Received: 16 February 2017; Accepted: 27 September 2017

\vspace*{1.0 cm}

accuracy assessments

\vspace*{1.0 cm}

\hrule

\vspace{0.5 cm}

\end{flushleft}

\clearpage

%{\bf Abstract} 

\begin{abstract}

The analysis and modelling of  a range of plasmas (for example:  astrophysical, laser-produced and fusion), require atomic data  for a number of parameters, such as energy levels, radiative rates and electron impact excitation rates, or equivalently the effective collision strengths. Such data are desired for a wide range of elements  and their many ions, although all elements are not useful for all types of plasmas. Since  measurements of atomic data are mostly confined to only a few energy levels of some ions, calculations for all parameters are highly important. However, often there are large discrepancies among different calculations for almost all parameters, which makes it difficult to apply the data with confidence. Many such discrepancies (and the possible remedies) were discussed earlier ({\em Fusion Sci. Tech.} {\bf 2013}, 63, 363). Since then a lot more anomalies for almost all of these atomic parameters have come to notice. Therefore, this paper is a revisit of various atomic parameters to highlight the large discrepancies, their possible sources  and some suggestions to avoid those,  so that comparatively more accurate and reliable atomic data may be available in the future.

\end{abstract}

\clearpage

\section{Introduction}
For the analysis and modelling of plasmas, atomic data for various parameters (including energy levels, radiative rates and electron impact excitation rates, or the effective collision strengths) are required. Measurements of these parameters  are mostly confined to only a few energy levels of some ions, and the compiled and assessed results are freely available on the NIST (National Institute of Standards and Technology) website: {\tt http://www.nist.gov/pml/data/asd.cfm}. Therefore, due to the paucity of experimental results,   calculations for all parameters are highly important. During the past few decades a wide range of methods and codes have been developed to calculate the desired data -- see for example \cite{fst} and references therein. However, often there are large discrepancies among different calculations for almost all parameters, as has already been discussed in detail earlier \cite{fst}. Such discrepancies  make it difficult for modellers to apply the data with confidence, particularly when neither measurements are available to compare with nor the calculations can be easily repeated to verify the accuracy.  Recently,  we have noticed that  anomalies for almost all of the atomic parameters persist in the literature. Therefore, the {\em aim} of this paper is to revisit the various atomic parameters to highlight the large discrepancies and their possible sources,  and to offer some suggestions (again) to avoid these,  so that comparatively more accurate and reliable atomic data may be available in the future. We also note here that several efforts have been made in the past by organising workshops on the reliability and accuracy of the atomic data, and suggestions have been made particularly about the 'uncertainty quantifications' of the data -- see for example, the recent paper by Chung et al. \cite{iaea}. However, our emphasis is more on the `large discrepancies', which may  propagate through the various stages of  a calculation, e.g. from structure to collisions, in ways that are  hard to quantify, as noted by \cite{iaea}.

Generally, for applications in astrophysical plasmas (to determine temperature, density and  chemical abundance) data are required for ions of lighter elements with Z $\le$ 30, whereas (particularly) for fusion plasmas  similar data  are also desired for heavier elements, such as Kr, Xe and W. In particular, tungsten is very important for the developing ITER project as it radiates at almost all ionisation stages. Additionally, atomic data are useful to model for the power loss due to the various impurities present in the reactor walls as well as to control the power generation by a deliberate introduction of these. Since the various atomic parameters have already been defined in our earlier paper \cite{fst}, we straightaway focus on the problems (discrepancies) noted in their determination. We also note here that most of the methods and codes applied in their determination have also been listed in the earlier paper and therefore are not repeated here.

\section {Energy Levels}

Calculations for energy levels are generally the easiest to perform and for this reason the literature is full of data for many ions of interest. Availability of measurements for some levels, as stated above, considerably helps in determining the theoretical energies and/or in improving the calculations, because various methodologies (such as optimising orbitals on particular state/level or the use of non-orthogonal or pseudo orbitals) can be applied. However, the problem becomes more difficult when no measurements (and even prior theoretical results) are available for comparisons, as recently experienced for three Cr-like ions: Kr~XIII, Tc~XX and Xe~XXXI \cite{crlike3}.  For many ions 'configuration interaction' (CI) among a few hundred levels (or configuration state functions, i.e. CSF) is sufficient to generate reasonably accurate energies  (within a few percents), but for some  much larger calculations ($\sim$10$^9$ or more CSFs) are necessary, such as for Cr-like ions \cite{crlike3, crlike1, crlike2}. 

Usually, discrepancies between theory and measurements (or among various calculations) are understandable, and can be resolved by different methodologies as stated above. For example, Turkington et al. \cite{turk} reported two sets of theoretical energies for the  levels of Na-like W~LXIV, for which they used two different atomic structure codes, namely CIV3 and GRASP.  Their CIV3 energies differ with those of NIST by up to 4~Ryd for a few levels, because the code does not include two-body relativistic operators  and quantum electro-dynamic effects (QED), which are very important for this heavy ion. However, their GRASP energies are comparatively much closer to the measurements, within $\sim$1~Ryd. These calculations have been performed with a different purpose in mind, mainly to compare the subsequent scattering data, and therefore the reasons for differences are understandable. However,  in a few instances it may not be straightforward. For example, Singh et al. \cite{mmbr} reported energy levels for 5 Br-like ions with 38 $\le$ Z $\le$ 42, for which they adopted the GRASP code with limited CI. However, in a subsequent calculation \cite{brlike1} with the same code and CI, it was noted that the calculations of Singh et al. cannot be reproduced for any of the ions considered by them.  Although the differences noted were only up to 0.15~Ryd ($\sim$5\%), it undermines the confidence in a calculation -- see also section~3 for much larger discrepancies noted for other parameters.

In any large calculation, particularly when the mixing among levels of the same and/or different configuration(s)  is strong, it is very difficult to assign  unambiguous designations for all levels --    see for example levels of Cl-like Fe~IX \cite{fe9}. This is a general atomic structure problem irrespective of the code adopted, and the best one can do is to provide the mixing coefficients so that the users can assign the designation according to their own preferences. It also helps to make comparisons with other calculations.  However, in general there are no discrepancies for the $J$ values. Similarly, in fully-relativistic calculations quantum labels for the targeted states are obtained in jj-coupling, and have a different mixing than in the semi-relativistic $LSJ$ work. Therefore, it often becomes difficult to match a correspondence between the two different labeling schemes, although the GRASP0 code of P.~H.~Norrington and I.~P.~Grant ({\tt http://amdpp.phys.strath.ac.uk/UK\_APAP/codes.html}) has this facility. Furthermore, recently a method has been developed by Gaigalas et al.  \cite{gag}, to transform the representation from jj to an approximate  LSJ-coupling, and to assign a unique label to all energy levels. 

It is not always possible to verify (or determine) the accuracy of energy levels by performing two (or more) independent calculations with different methods/codes. However, it is always beneficial to have some measurements to compare with and to improve the accuracy of theoretical results  -- see for example the levels of Si~II \cite{sstsi2} for which the use of non-orthogonal orbitals has been helpful. Another example for which the availability of measurements for a few levels has been very helpful is the case of Cr-like ions \cite{crlike1, crlike2}. As may be seen in table~A of \cite{crlike1} differences between the theoretical and experimental energies for the levels of Co~IV are up to 35\%, and the orderings are also different. This is in spite of using very large CI (with up to 76~138 CSFs), and with two different codes, namely GRASP and FAC.  For this (and other similar) ion(s) it became necessary to  consider over a 1000 configurations, for both the even and odd levels, which generated more than a billion CSFs. However, in any atomic structure calculation (code) it becomes difficult to handle such a large number of CSFs and therefore, we adopted the quasi-relativistic Hartree-Fock (QRHF) code of Bogdanovich and Rancova \cite{qr1, qr2}, in which  a reduction process was applied to scan and select the most important ones, based on their weights. This process reduced the number of CSFs to under a million for both the even and odd levels, but brought the discrepancies between theory and measurements below 3\% for the levels of the 3d$^6$ configuration, and below 1\% for the higher ones, i.e. (3d$^5$) 4s and 4p. 

 Another code which has been successfully applied with very large CI is the revised version of GRASP known as GRASP2K \cite{grasp2k, grasp2kk} -- see the work of J\"{o}nsson et al. \cite{blike1} for B-like ions,  of Froese Fischer \cite{cffw40} for W~XL, and the recent work of Gustafson et al. \cite{gustaf} for Mg-like iron. In all such cases the theoretical results are very close to the available measurements. This reinforces our earlier conclusion \cite{fst} that all codes are not useful for all applications, significant discrepancies between theory and measurements should not be ignored,  and  the choice of a code  is very important and should be based on: (i) the  ion concerned, (ii) the accuracy desired, and (iii) the application in mind. Furthermore, what configurations to include in a calculation  is equally important, and should be based on a fair balance of even and odd CSFs as deliberated by Froese Fischer \cite{cffw40}, and their energy ranges as emphasised by us \cite{w40a, w58a}.

\section{Radiative Rates and Lifetimes}

Unlike energy levels, it is comparatively more difficult to assess the accuracy of radiative rates (A-values), or the related parameters such as oscillator (f-values) and line (S-values) strengths -- see eqs (1)--(5) of \cite{ti19}. This is because experimental results for A-values are  very limited and are mostly determined through the measured lifetimes by combining the branching fractions -- see for example, the work of B\"{a}ckstr\"{o}m et al. \cite{back}. Therefore one has to (mostly)  rely on theoretical results. Nevertheless, in principle, all methods/codes should produce comparable results within a reasonable accuracy (say $\sim$20\%), provided the same level of CI is adopted. Unfortunately, that is not the case in some instances.

As stated earlier in section~2, Singh et al. \cite{mmbr} reported A-values for transitions in 5 Br-like ions with 38 $\le$ Z $\le$ 42,  adopting the GRASP code. However, their results could not be reproduced with the same version of the GRASP code and with the same configurations adopted by them, and their results differed from our work [8] by up to five orders of magnitude, and for all ions. The calculations are simple as only limited CI was used. The easiest way to understand such large discrepancies is to compare the input data between their  and our calculations, but we were not able to obtain the data from the authors.
Similarly, discrepancies in the reported A-values for W~XL and W~LVIII by S.~Aggarwal et al. \cite{mmw40} and Mohan et al. \cite{mmw58}, respectively, are up to four orders of magnitude for several transitions, as demonstrated by us \cite{w40a, w58a}. Furthermore,  discrepancies of  up to three orders of magnitude in A-values have been observed for several  transitions of Li-like ions  -- see Aggarwal and Keenan \cite{lilikea}. Therefore, it is always useful  to have more than one calculation for any ion in order to have confidence in the applied data.

Some discrepancies in A-values are unavoidable, particularly for weaker transitions (f $<$ 0.01), and when different codes and/or CI are used, and this has been discussed and demonstrated by various workers -- see for example fig.~1  of \cite{nrbelike}  for Be-like ions,  fig.~2 of \cite{sstmg5} for Mg~V and more recently fig.~1 of \cite{kbmgv} for Mg~V. This is because weaker transitions are more sensitive to the additive or cancellation effects of mixing coefficients from different sets  of CSFs and/or their energies ($\Delta E_{ij}$). Similarly, the ratio (R) of  the length and velocity forms of A-values may also  differ substantially as shown in fig. 1 of \cite{sstmg5}. However, the A- (or f-) values of comparatively strong(er) transitions  are (generally) more stable, and their values of R do not deviate much from  unity, which is an indication of accuracy --  see also \cite{uq}. Nevertheless, this is only a desirable criterion and not a necessary one because even for very strong transitions different calculations may yield R close to unity but strikingly different f-values as demonstrated in our work on Mg-like ions \cite{mglike} and on Ne-like Ni~XIX \cite{ni19}. Most of these differences are normally understandable and often do not exceed over an order of magnitude.

As already stated above, direct measurements of A-values are  very limited in the literature. However, lifetimes ($\tau$ = 1.0/${\sum_{i}} A_{ji}$, s) for several levels of many ions have been measured by different techniques, and hence provide an indirect comparison (and thus assessment of accuracy) of the A-values. Generally, A-values for electric dipole (E1) transitions are dominant and hence the most important. However, similar data are also required for electric quadrupole (E2), magnetic dipole (M1) and  magnetic quadrupole (M2) transitions.  A-values for all types of transitions improve the accuracy of the plasma model. Additionally, if the A-value for any (type of) transition dominates in the determination of $\tau$ for any level then one gets the 'experimental' A-value.  Often, as among various calculations, significant differences (of over an order of magnitude) are also noted among experimental results -- see for example, table~5 of \cite{ciii} for the levels of C~III and table~5 of \cite{niv} for N~IV. If several measurements over a period of time and by different techniques are available, as in the case of above two ions, then differences with theory can be ironed out. However, problem arises when a single set of measurements are available, as  for the levels of Si~II \cite{si2a}.   For the 3s3p$^2$ $^4$P$_{1/2,3/2,5/2}$ levels  of Si~II,  Calamai et al. \cite{cal} have measured $\tau$ which are lower than our theoretical results \cite{si2a} by an order of magnitude. The dominant contributing E1  transitions for these levels are invariably weak as their  f- values are $\sim$10$^{-7}$, and hence large variations among different calculations are not uncommon as discussed above and also seen in table~3 and fig.~3 of \cite{baut1}. 

Unfortunately, Si~II is a difficult ion because even with various options of CI, orbitals, methodology, etc. differences between theoretical and experimental energy levels remain significant -- see for example, table~2 of \cite{baut1} and table~1 of \cite{sstsi2a}. Since measurements of $\tau$ for a few levels of Si~II are available, as stated above, it is a general practice in any theoretical work to match the calculations with them, and this is possible by adopting different combinations of CSFs and/or the use of non orthogonal orbitals. As a result, the A-values for the concerned transitions agree closely between some theoretical work (for example \cite{sstsi2a}) and the measurements of \cite{cal}. However, adoption of this data in the analysis of observations from astrophysical plasmas led to very large discrepancies and prompted Baldwin et al. \cite{bald} to pronounce it as a ``Si ~II disaster ...". On the other hand, adoption of our (simple) A-values \cite{si2a} resulted in a much better match between theory and observations, and resolved some of the discrepancies, as discussed in detail by Laha et al. \cite{si2b, si2c}. Therefore, the measurements of Calamai et al. \cite{cal} may or may not be accurate, and another set of experimental results of $\tau$ for the levels of Si~II will be highly useful for both: (i) the assessment of accuracy of theoretical results, and (ii) further analysis of astrophysical observations. Similarly, a much larger ({\em ab initio}) calculation for Si~II will be highly useful to settle these discrepancies.

The discrepancies in $\tau$ noted above for the levels of a few ions, either between theory and measurements or among various theoretical results, are comparatively minor and mostly understandable. However, there are other examples for which discrepancies in $\tau$ are much larger and are not understandable. For example, for the levels of Br-like W~XL the discrepancies between the A-values of \cite{mmw40} and our work \cite{w40a} were only up to two orders of magnitude, but much larger (up to four orders of magnitude) for $\tau$. Similar discrepancies and of the same orders of magnitude have also been noted for the levels of W~LVIII -- see \cite{w58a} and \cite{mmw58}. On the other hand, there are no discrepancies between the A-values of \cite{mmw62} and our work \cite{w62a}, but the $\tau$ results differ by up to 14 orders of magnitude for over 90\% levels of W~LXII. This was due to random selection of A-values by \cite{mmw62} in the determination of $\tau$ as explained by us \cite{w62a}. Finally, similar discrepancies and for the same randomness reason have been noted for the levels of W~LXVI -- see \cite{mmw66} and \cite{w66b}. In conclusion, such   situations create unnecessary problems for the users and/or assessors of atomic data as without performing an independent calculation  results cannot be taken at the face value. Therefore as for energy levels, multiple calculations for A-values and preferably by differing methodologies are always useful to resolve large discrepancies and to have confidence in the applied data.  This point has also been emphasised by Chung et al. \cite{iaea}, and has most successfully been implemented by Wang et al. \cite{wang1, wang2, wang3} and Si et al. \cite{si} for a series of ions. They have performed systematic calculations with the MCDF and MBPT methods to determine the energy levels and A-values to spectroscopic accuracy. However, for this kind of work very large calculations need to be performed for which J\"{o}nsson et al. \cite{alex} have suggested further modifications to the GRASP2K code. Similarly, Safronova et al. \cite{saf1, saf2} have performed very accurate (and large) calculations for a series of ions, using the CI+MBPT codes, and with single, double and triple excitations of electrons. 

\section{Collision Strengths and Effective Collision Strengths}

The collision strength ($\Omega$) is a dimensionless parameter and is very simply related to the scattering cross section ($\sigma$) -- see eq. (4) of \cite{fst}. Measurements of $\sigma$ are restricted to only a couple of transitions, in a limited energy range, and for only some ions. Whenever measurements for $\sigma$ are available, generally there is a good agreement with the corresponding theoretical work/s -- see for example,  figs. 2 and 3 of \cite{wall},  fig. 4 of \cite{rise}, and  fig. 1 of \cite{kai} for the 3s$^2$~$^1$S -- 3s3p~$^{1,3}$P$^o$ transitions of Si~III. Such agreements between theory and measurements are helpful for assessing the validity and accuracy of theory, but are not very useful for applications for which much larger range of data are required. Hence, a vast amount of data available in the literature for $\Omega$ are theoretical, obtained with a variety of scattering codes (mostly distorted-wave DW and $R$-matrix). Furthermore, such data cover a much wider range of energy (up to several times the thresholds) and  transitions, involving up to $\sim$500  levels, or even more. 

As for A-values, often differences among several calculations for $\Omega$ are also large, and this is in spite of adopting the same level of complexity, or even methodology. Several examples (and reasons) of large discrepancies  for some ions were  discussed in our earlier paper \cite{fst}, and hence are not repeated here. Furthermore, it is the {\em effective} collision strength ($\Upsilon$) for which data are useful for applications, rather  than for $\Omega$. This is because often plasmas have an electron energy distribution and therefore values of $\Omega$  need to be averaged over a suitable distribution, most often {\em Maxwellian}, to obtain results for $\Upsilon$ -- see eq. (5) of \cite{fst}. Therefore, we focus our attention mostly on the data for $\Upsilon$. Additionally, even large discrepancies in $\Omega$ values, at some energies,  may not necessarily lead to corresponding differences in the $\Upsilon$ values, because the averaging of the former is performed over a wide energy range, and the values of the latter are highly dependent on the temperature and/or the (type of) transition. As an example, see table~D of \cite{si3} for comparison of $\Omega$s and table~E for $\Upsilon$s, for transitions of Si~III. Finally, we also note here that the values of $\Omega$ vary smoothly at energies {\em above} thresholds, and below that show highly dense and  numerous (closed channel or Feshbach) resonances -- see for example, fig.~3 of \cite{mgv} for three transitions of Mg~V.

Differences of up to a factor of two or so in values of $\Upsilon$ between any two calculations are not uncommon, and this is particularly true at low(er) temperatures for which a slight variation in the position and/or  magnitude of resonances can have large consequences on the values of $\Upsilon$. For (particularly strong) allowed transitions the contribution of resonances, if any, is not significant and therefore it is much easier to understand differences in $\Upsilon$ between any two calculations, as the results simply follow the f-values, particularly towards the high(er) end of the temperature range. On the other hand, forbidden transitions are full of resonances and hence the large variations noted in their $\Upsilon$ values. Nevertheless, problems arise for the users and modellers of data when large discrepancies, of orders of magnitude, are noted for $\Upsilon$ between any two independent calculations, and that too with the same (or similar) method and complexity. Below we discuss such discrepancies in detail. 

The calculations of $\Omega$ (and subsequently $\Upsilon$) performed with the DW code(s) do not (generally) include the contribution of resonances, and therefore their $\Upsilon$ values are expectedly underestimated, particularly for the forbidden transitions. A widely used and highly successful method which explicitly includes the contribution of resonances is the $R$-matrix, of which several versions with some variations are available, as noted and referenced in \cite{fst}. For most ions for astrophysical applications (Z $\le$ 28) the non-relativistic version is  generally sufficient to achieve a reasonable accuracy, although a fully relativistic version (DARC: Dirac atomic R-matrix code) is preferable. This is because it includes {\em fine structure} in the definition of channel coupling and therefore resonances arising in between the degenerate levels of a state can also be included.  Similarly, for some ions, such as He~II \cite{he2}, the $LS$ states are non-degenerate in energies, but the levels split with the inclusion of QED. However, precisely for the same reason the size of the Hamiltonian (H) matrix increases substantially, and therefore the calculations become more time and resource consuming, without giving any overall significant advantage. Nevertheless, results of $\Upsilon$ for several ions are available in the literature from both DARC and the non-relativistic version. However, calculations with the latter are primarily in the $LS$ (Russell-Saunders or spin-orbit) coupling, and the subsequent results for $\Omega$ (and $\Upsilon$) for  fine-structure transitions are calculated through the ICFT (intermediate coupling frame transformation) approach, as described in  \cite{icft}. Unfortunately, large discrepancies between the DARC and ICFT results for $\Upsilon$ have been noted for a wide range of ions, and on several occasions. We elaborate on this below.

Discrepancies in the $\Upsilon$ results, obtained with the DARC and ICFT codes, until 2012 have mostly been discussed in our earlier paper \cite{fst}, and therefore these are not repeated here. However, just to recapitulate large discrepancies had been noted (and highlighted) for H-like, He-like, Li-like, and some Mg-like ions -- see \cite{fst} for problems and specific references. Here we mostly focus on transitions in Be-like (and some other specific) ions which have been under discussion in the recent literature. To be specific, in the recent past we have reported results with DARC for 7 Be-like ions, namely  C~III, N~IV, Al~X, Cl~XIV, K~XVI, Ti~XIX, and Ge~XXIX \cite{ti19, ciii, niv, belike, alx1},    and similar data for a wide range of Be-like ions with Z $\le$ 36 have been reported in \cite{nrbelike}. Although most of the discrepancies, of orders of magnitude, have already been highlighted and discussed in some of our papers (see \cite{ciii, niv, alx2}), we elaborate below on the (possible) reasons for these.

If any two (or more)  calculations are performed with the same method and are of comparable complexity then the results for $\Upsilon$ (and other related parameters) should be similar for most of the transitions (i.e. the differences should not be very large), and particularly at temperatures towards the higher end at which the contributions of resonances, if any, are expected to be insignificant. However, that is not the case for Be-like ions, as stated above. One of the reasons for large discrepancies could be a significant difference in  the level of complexity of the calculations. For example, our calculations for Al~X, Cl~XIV, K~XVI, Ti~XIX, and Ge~XXIX \cite{ti19,  belike, alx1} included only 98 levels of the 2$\ell$2$\ell'$, 2$\ell$3$\ell$ and  2$\ell$4$\ell$  configurations, whereas those of \cite{nrbelike} were larger with 238 levels, the additional 140 belonging to  2$\ell$5$\ell$, 2s6s/p/d, 2p6s/p/d, 2s7s/p/d, and 2p7s/p/d. Therefore, the discrepancies of over an order of magnitude in $\Upsilon$ results,  observed for about 50\% of the transitions and at all temperatures, may be because of the larger model adopted by \cite{nrbelike}. Since in a majority of cases $\Upsilon_{ICFT}$ were greater than $\Upsilon_{DARC}$  -- particularly see fig. 2 of \cite{alx2} -- it appears to be a sound reason, as discussed and demonstrated  in a series of papers by Badnell and co-workers \cite{icft2, icft3, icft4}. Below we elaborate more on the differences between the DARC and ICFT calculations.

In Figs. 1--3 (a and b) we show the differences between $\Upsilon_{DARC}$ and $\Upsilon_{ICFT}$ in the form of ratio (R) for transitions in Al~X, C~III and N~IV, at a single  temperature of 1.0$\times$10$^6$, 9.0$\times$10$^4$ and 1.6$\times$10$^5$~K, respectively. These temperatures are the most relevant (and common between the two calculations) for these respective ions. Additionally, progessively larger calculations with DARC have been performed for these ions, i.e. Al~X (98 levels \cite{ti19}), C~III (166 levels \cite{ciii}) and N~IV (238 levels \cite{niv}), whereas the ICFT results \cite{nrbelike} for all ions include 238 levels of the above listed 39 configurations. Additionally, for each ion two panels are shown, one (a) for transitions {\em from} the lower  (I) and the other (b) {\em to} upper levels (J). This is because sometimes they reveal different (i.e. more specific) conclusions. Finally, for brevity and clarity only those transitions are shown which differ by over 20\%.

For Al~X comparisons in Fig.~1 are shown only among the lowest 78 levels because beyond that those from higher configurations (included in ICFT but not in DARC) intermix. It is clear from this figure that for a majority of transitions $\Upsilon_{\rm ICFT}$ $>$ $\Upsilon_{\rm DARC}$, and a reasonable agreement between the two independent calculations is only for transitions among the lowest $\sim$60 levels, as clearly seen in Fig.~1(b). For this ion the reasoning of a larger model given by \cite{icft2} and shown in their fig.~2 appears to be plausible. However, even for the transitions in C~III for which our calculations \cite{ciii} included a considerably larger number of levels (166), the discrepancies are large as shown in Fig.~2. In fact, for transitions only among the lowest 20 levels there is a satisfactory agreement between the two calculations -- see Fig.~2(b).   Therefore, the differences seen here do not appear to be  because of a larger model included by \cite{nrbelike}. This is further confirmed by Fig.~3 for the transitions of N~IV for which both calculations include exactly the same number of levels (and from the same configurations), but the satisfactory agreement is only for transitions among the lowest $\sim$30 levels. Furthermore, these differences (discrepancies) increase with increasing temperature as discussed and demonstrated in our earlier papers \cite{ciii, niv, alx2}. Similarly, the discrepancies shown here in Figs. 1--3 are only for a limited number of transitions. Considering all transitions these are much larger. For example, at T$_e$ = 1.6$\times$10$^{6}$~K discrepancies of over 20\% are up to four orders of magnitude for about 40\% of the transitions of N~IV \cite{niv}, and in a majority of cases $\Upsilon_{\rm ICFT}$ $>$ $\Upsilon_{\rm DARC}$. We also note here that another calculation with DARC was performed for N~IV with the same 166 levels as for C~III. However, differences between the two sets of $\Upsilon$ were (expectedly) insignificant at all temperatures, and thus further confirming that a larger model does not necessarily improve the calculated results. We discuss the reasons below for  large discrepancies noted between the DARC and ICFT results.

\begin{figure*}
% \vspace{250pt}
\includegraphics[angle=-90,width=0.9\textwidth]{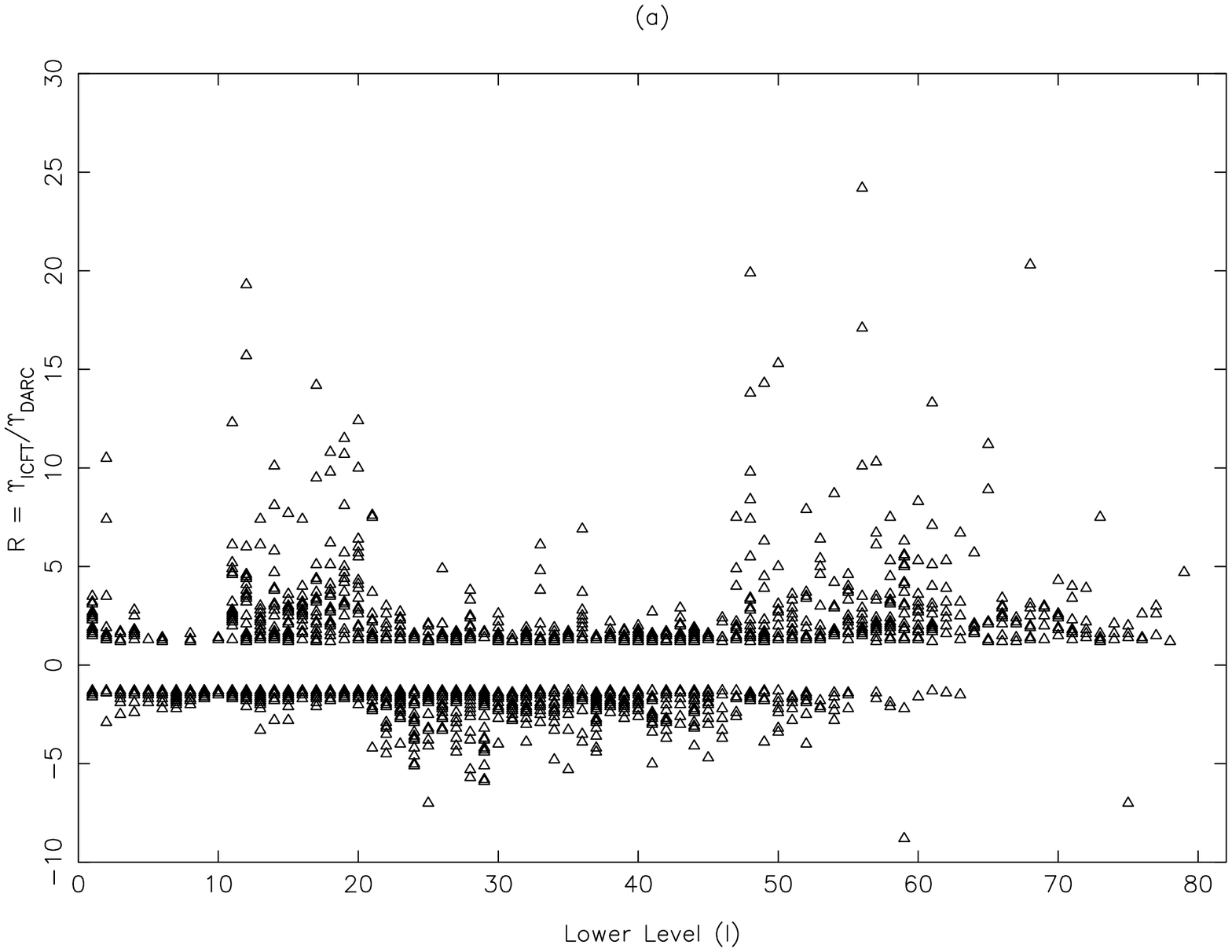}
 \vspace{-1.5cm}
% \caption{Comparison of DARC and ICFT values of $\Upsilon$ for transitions of Al~X at  T$_e$ = 1.0$\times$10$^{6}$ K. Negative R values plot  $\Upsilon_{\rm DARC}$/$\Upsilon_{\rm ICFT}$ and indicate that $\Upsilon_{\rm DARC}$ $>$ $\Upsilon_{\rm ICFT}$. Only those transitions are shown which differ by over 20\%. (a) Transitions {\em from} lower and (b) {\em to} upper levels.}
 \end{figure*}
 
\begin{figure*}
% \vspace{250pt}
\includegraphics[angle=-90,width=0.9\textwidth]{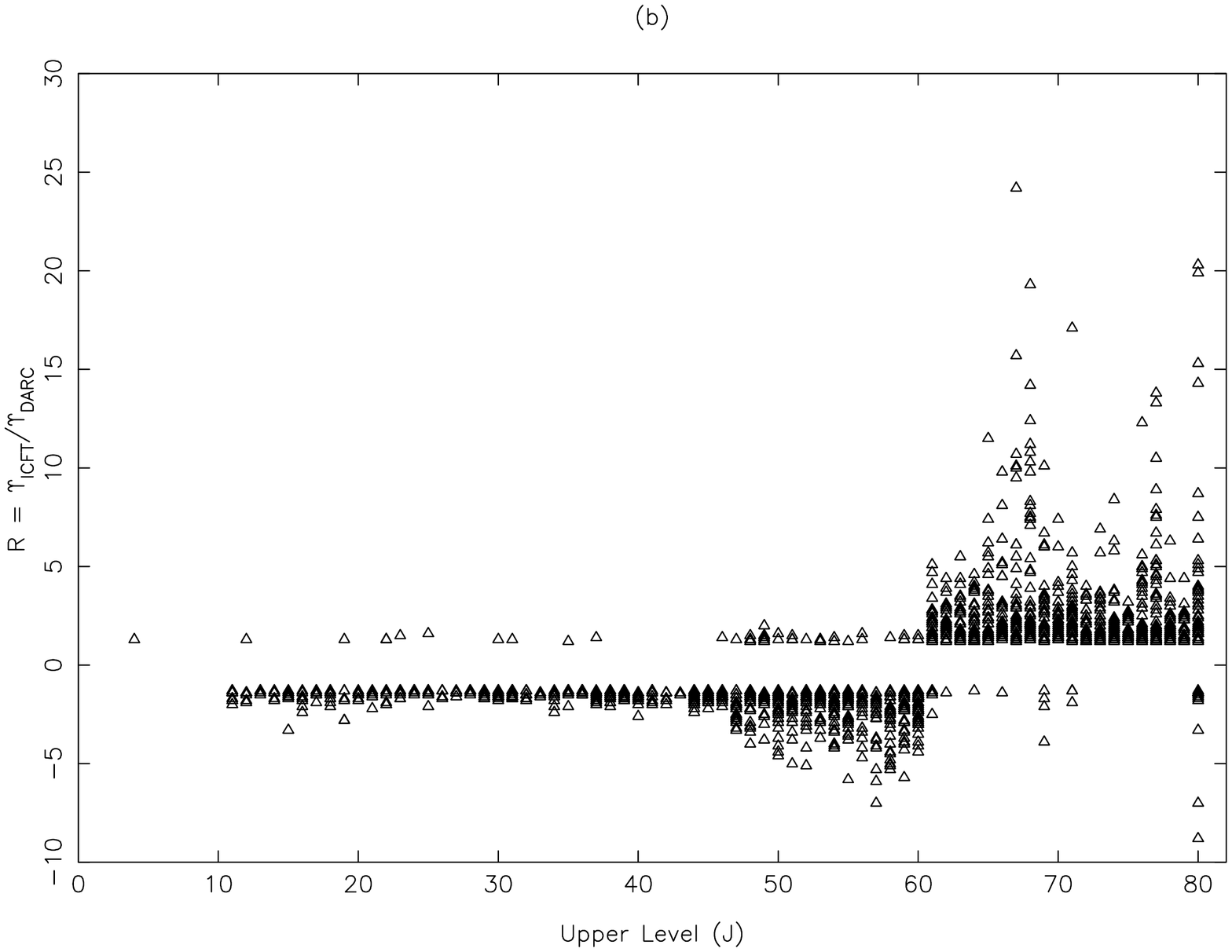}
 \vspace{-1.5cm}
 \caption{Comparison of DARC \cite{alx1} and ICFT \cite{nrbelike} values of $\Upsilon$ for transitions of Al~X at  T$_e$ = 1.0$\times$10$^{6}$~K. Negative R values plot  $\Upsilon_{\rm DARC}$/$\Upsilon_{\rm ICFT}$ and indicate that $\Upsilon_{\rm DARC}$ $>$ $\Upsilon_{\rm ICFT}$. Only those transitions are shown which differ by over 20\%. (a) Transitions {\em from} lower and (b) {\em to} upper levels.}
 \end{figure*}

\begin{figure*}
% \vspace{250pt}
\includegraphics[angle=-90,width=0.9\textwidth]{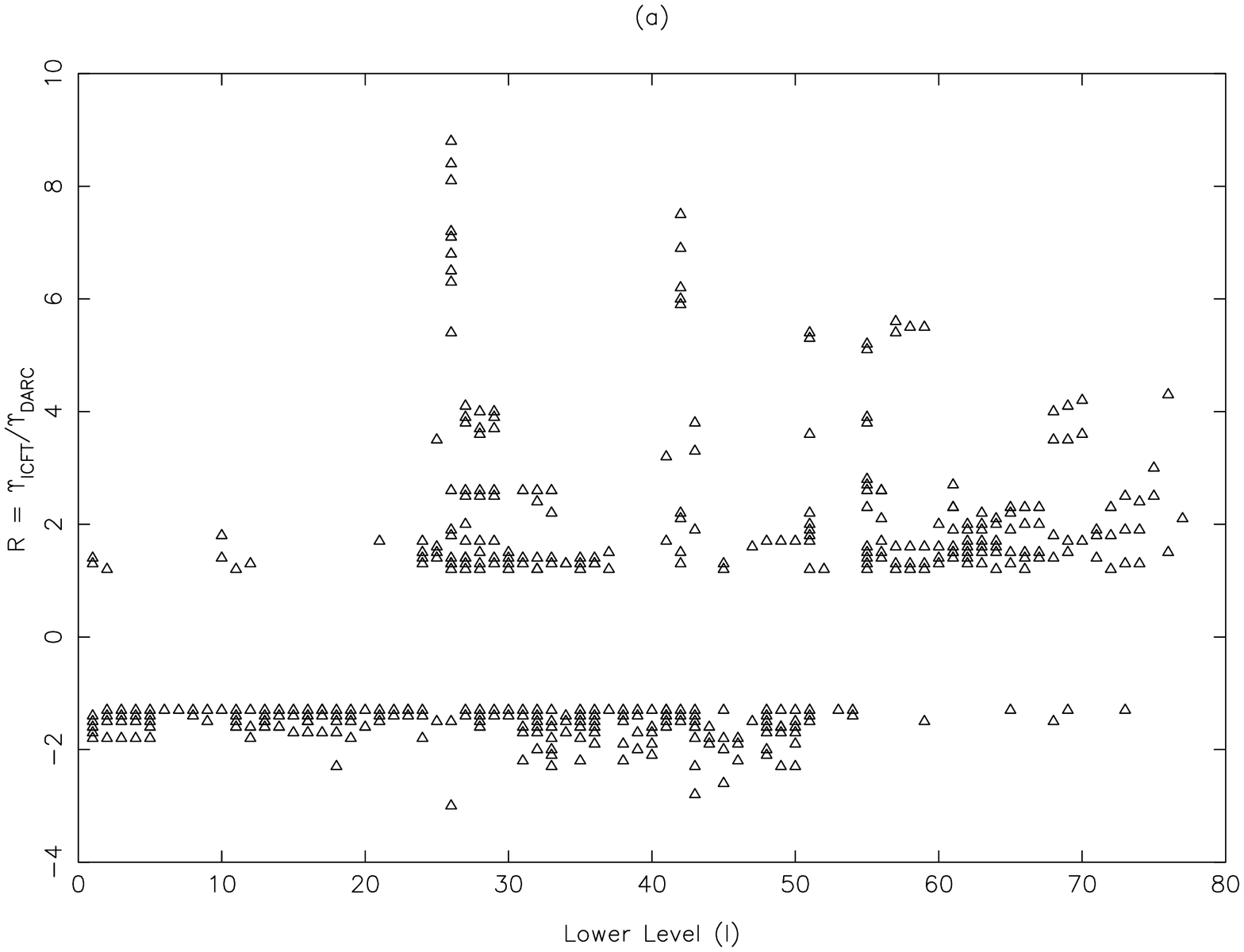}
 \vspace{-1.5cm}
% \caption{Comparison of DARC and ICFT values of $\Upsilon$ for transitions of C~III at  T$_e$ = 9.0$\times$10$^{4}$ K. Negative R values plot  $\Upsilon_{\rm DARC}$/$\Upsilon_{\rm ICFT}$ and indicate that $\Upsilon_{\rm DARC}$ $>$ $\Upsilon_{\rm ICFT}$. Only those transitions are shown which differ by over 20\%. (a) Transitions {\em from} lower and (b) {\em to} upper levels.}
 \end{figure*}
 
\begin{figure*}
% \vspace{250pt}
\includegraphics[angle=-90,width=0.9\textwidth]{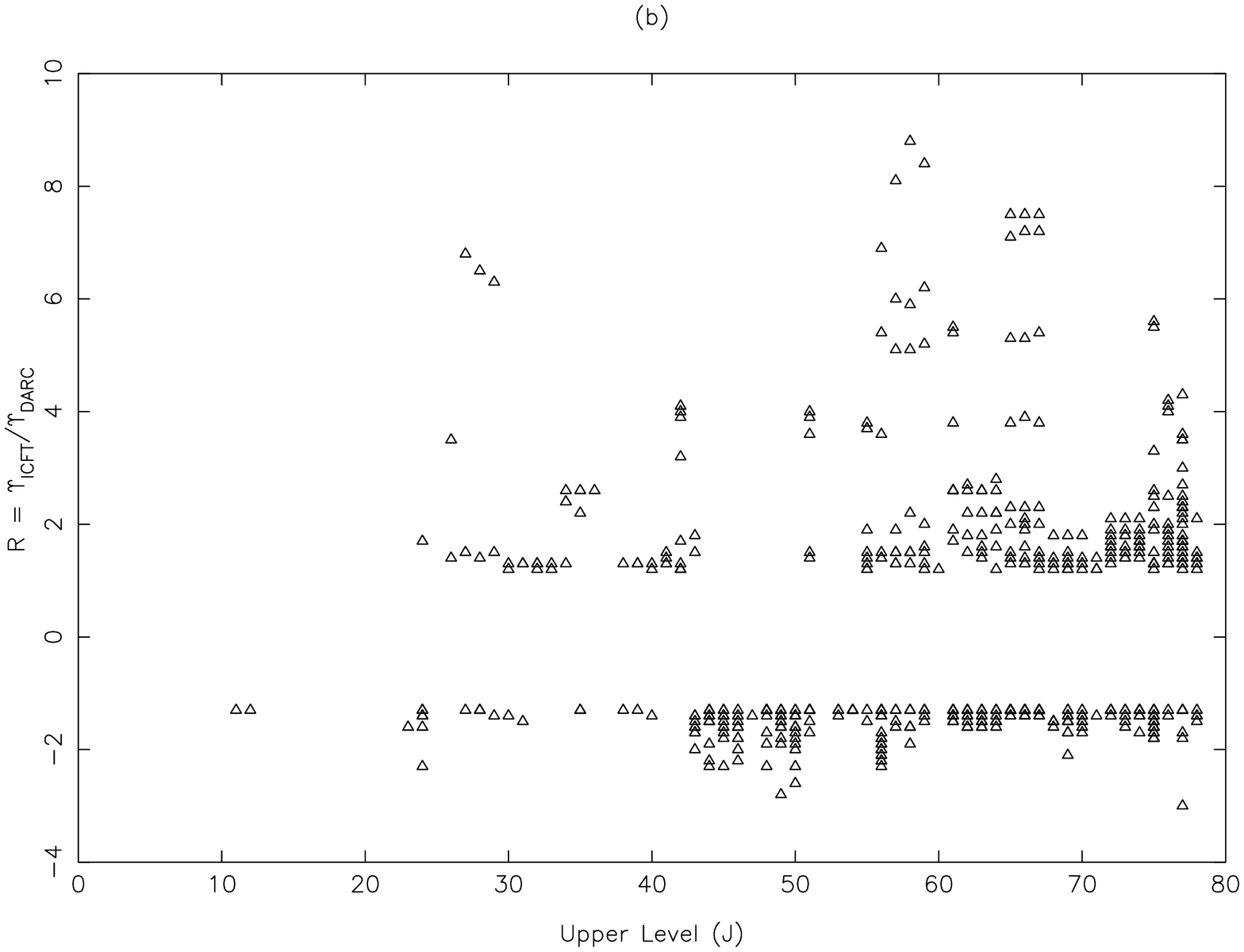}
 \vspace{-1.5cm}
 \caption{Comparison of DARC \cite{ciii} and ICFT \cite{nrbelike} values of $\Upsilon$ for transitions of C~III at  T$_e$ = 9.0$\times$10$^{4}$~K. Negative R values plot  $\Upsilon_{\rm DARC}$/$\Upsilon_{\rm ICFT}$ and indicate that $\Upsilon_{\rm DARC}$ $>$ $\Upsilon_{\rm ICFT}$. Only those transitions are shown which differ by over 20\%. (a) Transitions {\em from} lower and (b) {\em to} upper levels.}
 \end{figure*}

\begin{figure*}
% \vspace{250pt}
\includegraphics[angle=-90,width=0.9\textwidth]{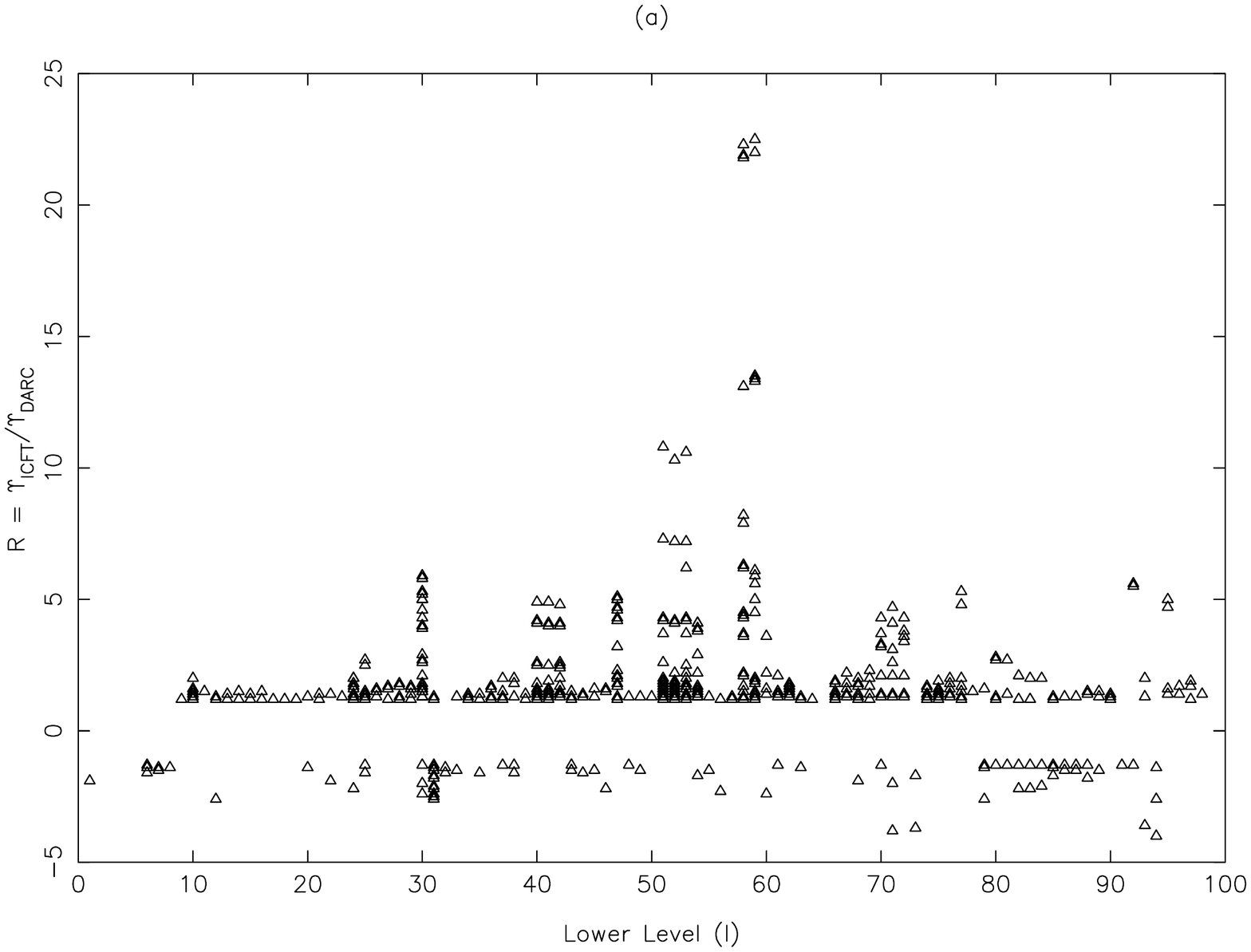}
 \vspace{-1.5cm}
% \caption{Comparison of DARC and ICFT values of $\Upsilon$ for transitions of N~IV at  T$_e$ = 1.6$\times$10$^{5}$ K. Negative R values plot  $\Upsilon_{\rm DARC}$/$\Upsilon_{\rm ICFT}$ and indicate that $\Upsilon_{\rm DARC}$ $>$ $\Upsilon_{\rm ICFT}$. Only those transitions are shown which differ by over 20\%. (a) Transitions {\em from} lower and (b) {\em to} upper levels.}
 \end{figure*}
 
\begin{figure*}
% \vspace{250pt}
\includegraphics[angle=-90,width=0.9\textwidth]{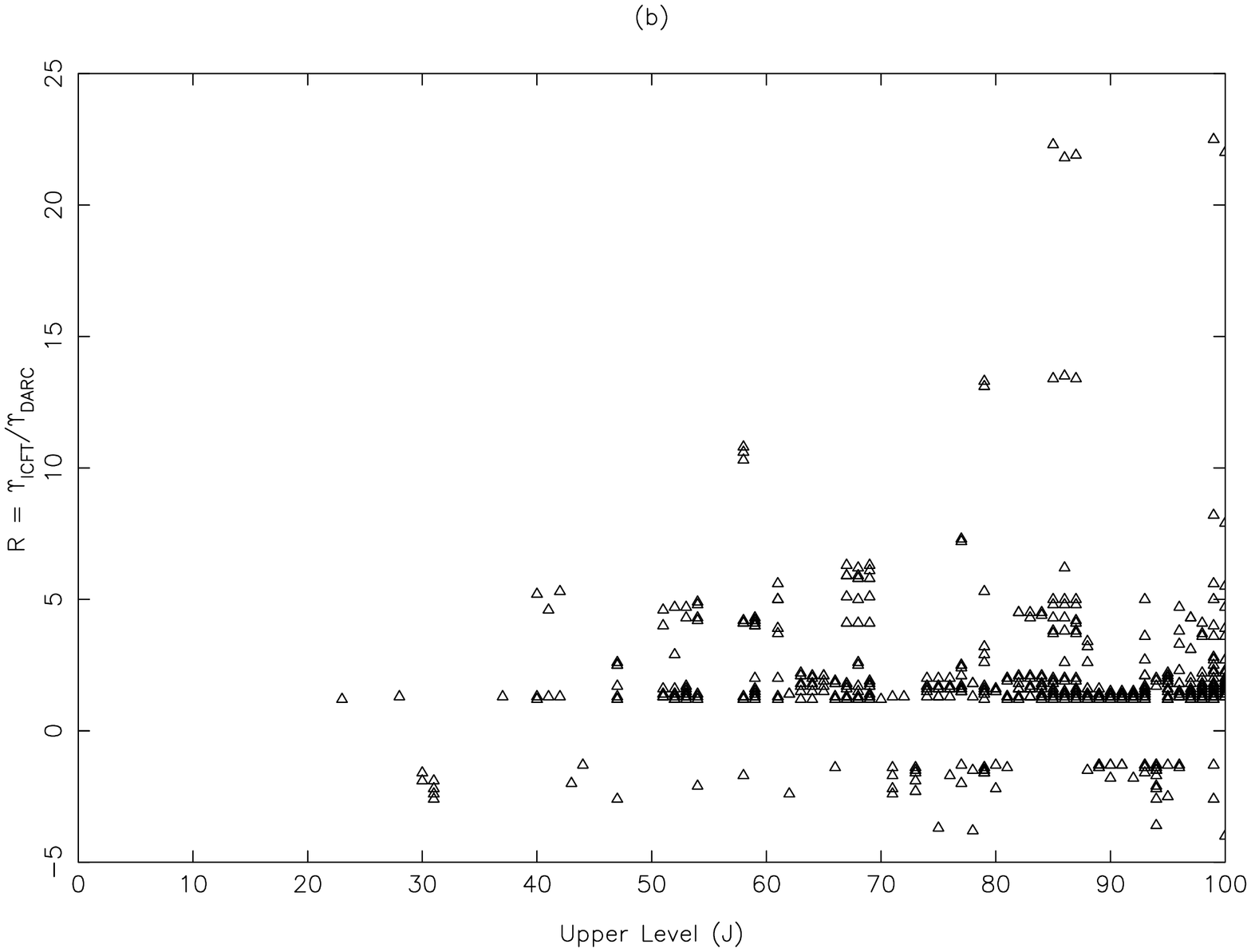}
 \vspace{-1.5cm}
 \caption{Comparison of DARC \cite{niv} and ICFT \cite{nrbelike} values of $\Upsilon$ for transitions of N~IV at  T$_e$ = 1.6$\times$10$^{5}$~K. Negative R values plot  $\Upsilon_{\rm DARC}$/$\Upsilon_{\rm ICFT}$ and indicate that $\Upsilon_{\rm DARC}$ $>$ $\Upsilon_{\rm ICFT}$. Only those transitions are shown which differ by over 20\%. (a) Transitions {\em from} lower and (b) {\em to} upper levels.}
 \end{figure*}

Apart from the included number of levels (discussed above) there can be several other reasons for the large discrepancies in $\Upsilon$, such as (i) (non) convergence of $\Omega$ w.r.t. partial waves, (ii) insufficient range of energy included for the determination of $\Upsilon$, (iii) energy resolution of resonances, (iv) presence of  pseudo and/or spurious resonances, and (v) differences in  the CSFs. However, none of these factors is responsible for the large discrepancies noted above, and already discussed in detail \cite{niv, alx2}. The likely reason is that the ICFT calculations for $\Omega$ are mostly performed up to a very limited energy range whereas the results for $\Upsilon$ are determined up to very high temperature. In order to do that the range of $\Omega$ is extended to high (infinite) energy through the formulations suggested by Burgess and Tully \cite{bt}, whereas the DARC calculations for $\Omega$ are performed up to reasonably  high energy to determine $\Upsilon$ up to a limited range of T$_e$, but sufficient for applications to a wide range of plasmas. For example, for N~IV  Fern{\'a}ndez-Menchero et al. \cite{nrbelike} calculated values of $\Omega$ up to an energy of 17~Ryd (or only 11~Ryd {\em above} thresholds) but reported $\Upsilon$ results up to 3.2$\times$10$^7$~K, equivalent to $\sim$200~Ryd. Therefore, they have {\em extrapolated} their $\Omega$ results over a very (very) high energy range and that is the likely cause of large discrepancies noted between the DARC and ICFT calculations. For the same reason the behaviour of their $\Upsilon$ results for several transitions is abnormal. For example, for the forbidden transitions their $\Upsilon$ values increase with increasing T$_e$ whereas the opposite is expected -- see \cite{niv,  alx2} for detailed comparisons and discussion. We also stress here that the problem is not with the ICFT methodology, but its implementation. If the calculations are performed with reasonable compromises then there should be no large discrepancies between any two calculations and this has been discussed and demonstrated on several earlier occasions (such as \cite{icft5}), and most recently  by Turkington et al. \cite{turk} for the transitions of W~LXIV -- note particularly the differences of up to 3~Ryd in energy levels  (as discussed in section~2) but similarity of results for $\Omega$ and $\Upsilon$ as shown in their  table~III and figs.~1--5.

Most of the discrepancies noted in $\Upsilon$ values for any transition between any two independent calculations are (often) a direct consequence of the corresponding differences in the $\Omega$ values. Unfortunately, a majority of authors do not report the $\Omega$ results, and hence it becomes difficult to identify the real reason for the large discrepancies. However, the lead author of \cite{nrbelike}, i.e. Luis Fern{\'a}ndez-Menchero, has very kindly supplied us the $\Omega$ data for a few transitions of N~IV, particularly those for which the discrepancies between the DARC and ICFT $\Upsilon$ are the greatest. In Figs.~4 and 5 we show their $\Omega$ in the 6--16~Ryd energy range for only two transitions, namely 30--232 (2s4s~$^1$S$_0$ -- 2p7d~$^3$D$^o_3$) and 30--235 (2s4s~$^1$S$_0$ -- 2p7d~$^3$P$^o_0$) -- see table~1 of \cite{niv} for the definition of all energy levels. Both these transitions are {\em forbidden}, but their upper levels are very close to the highest level 238 (2p7d~$^1$S$_0$) considered in both calculations. It means that the energy range for resonances (if any) is very narrow ($<$ 0.02~Ryd) and insignificant, and hence the variation of $\Upsilon$ should closely follow that for their $\Omega$ values. It is clear from these two figures (which have been shared with  Luis Fern{\'a}ndez-Menchero) that the behaviour of their $\Omega$ results is not correct, as these should considerably decrease with increasing energy (as is the case in our calculations although not shown here for brevity). Similar discrepancies and abnormal behaviour of $\Omega$ were noted earlier for transitions in Fe~XV -- see figs. 1--6 of \cite{fe15a}, but were later rectified by Berrington et al. \cite{icft5}.

\begin{figure*}
% \vspace{250pt}
\includegraphics[angle=90,width=0.9\textwidth]{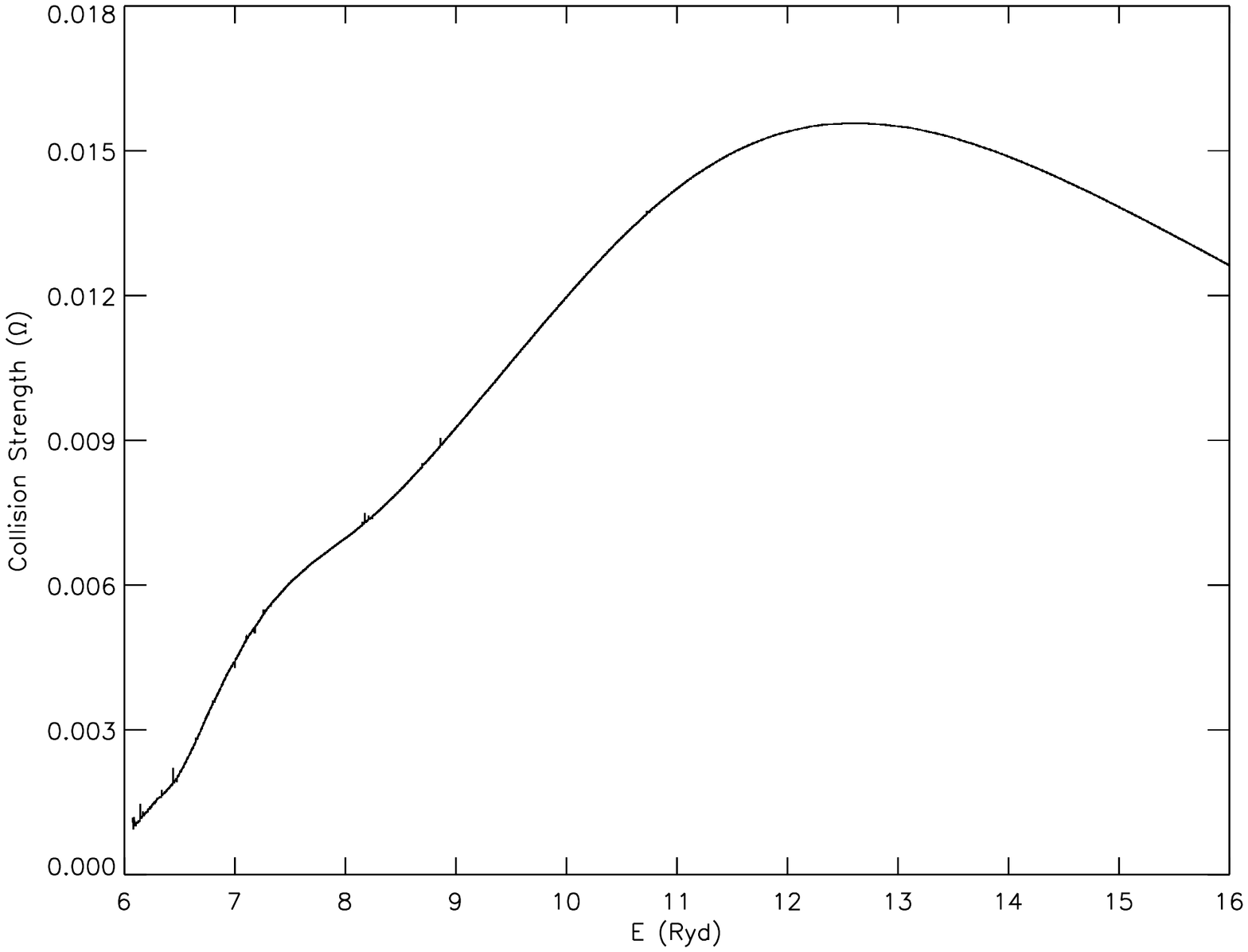}
 \vspace{-1.5cm}
 \caption{Collision strengths for the 30--232 (2s4s~$^1$S$_0$ -- 2p7d~$^3$D$^o_3$) transition of N~IV from the ICFT calculations of \cite{nrbelike}.}
 \end{figure*}

\begin{figure*}
% \vspace{250pt}
\includegraphics[angle=90,width=0.9\textwidth]{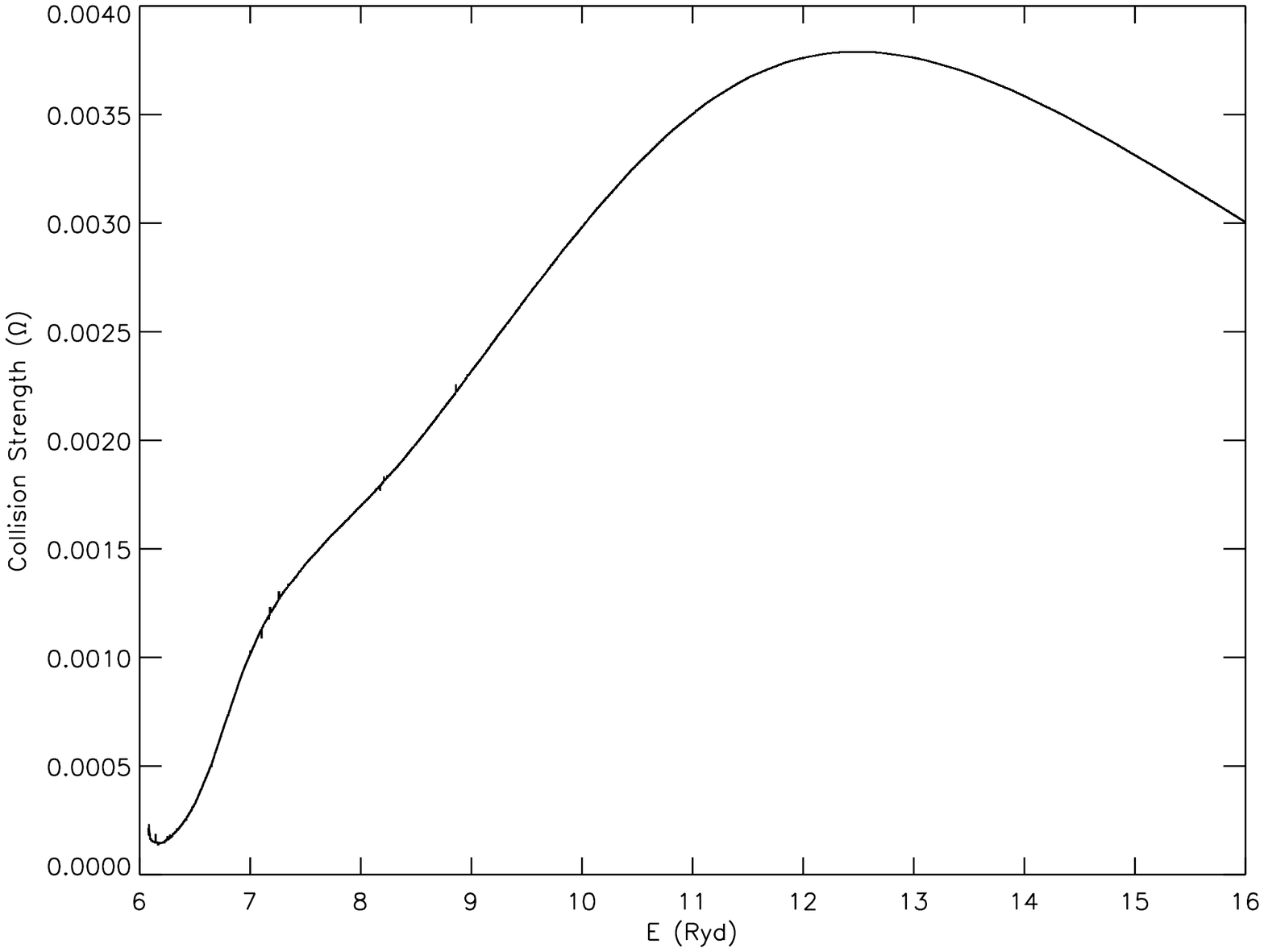}
 \vspace{-1.5cm}
 \caption{Collision strengths for the 30--235 (2s4s~$^1$S$_0$ -- 2p7d~$^3$P$^o_0$) transition of N~IV from the ICFT calculations of \cite{nrbelike}.}
 \end{figure*}

The overestimation of $\Upsilon$ values through the ICFT approach has not been restricted to the Be-like ions alone, but to others as well -- see \cite{fst} for H-like, He-like and Li-like ions. More recently, similar discrepancies (and overestimations of $\Upsilon$) have also been noted for Al-like Fe~XIV \cite{fe14} (although later defended by Del Zanna et al. \cite{icft3}) and Ar-like Fe~IX \cite{tz}. Finally, we briefly discuss transitions in Si~III, which is an important Mg-like ion for astrophysical applications. For Mg-like ions, including Si~III,  Fern{\'a}ndez-Menchero et al.  \cite{nrmglike} have performed ICFT calculations with 283 levels belonging to the 3$\ell3\ell'$, 3$\ell$4$\ell$ and 3$\ell$5$\ell$ configurations. On the other hand, our calculations with DARC \cite{si3} include only 141 levels of the 3$\ell3\ell'$ and  3$\ell$4$\ell$ configurations, but discrepancies between the two sets of $\Upsilon$ are over 20\% and up to two orders of magnitude for about half the transitions, and at all temperatures --  particularly see  fig.~5 of \cite{si3} for the anomalous behaviour of $\Upsilon$ by \cite{nrmglike}. These discrepancies have already been discussed in our recent paper \cite{si3}, but in Fig.~6 (a and b) we show the comparisons of transitions for Si~III at T$_e$ = 1.8$\times$10$^6$~K, as in Figs. 1--3 for Be-like ions. Unfortunately, the only transitions for which there is satisfactory agreement between the two calculations are those within the lowest 29 levels alone. It may be worth noting here that the additional 142 levels included in \cite{nrmglike} are not above the 141 considered in \cite{si3}, but are closely intermixed. In some instances the levels of a higher configuration are distinctly above those of the lower ones, and hence clearly influence the calculations for $\Upsilon$, particularly for  transitions among the lower levels -- see for example the case of Mo~XXXIV \cite{mo34} or neutral helium \cite{bk}. 

\begin{figure*}
% \vspace{250pt}
\includegraphics[angle=-90,width=0.9\textwidth]{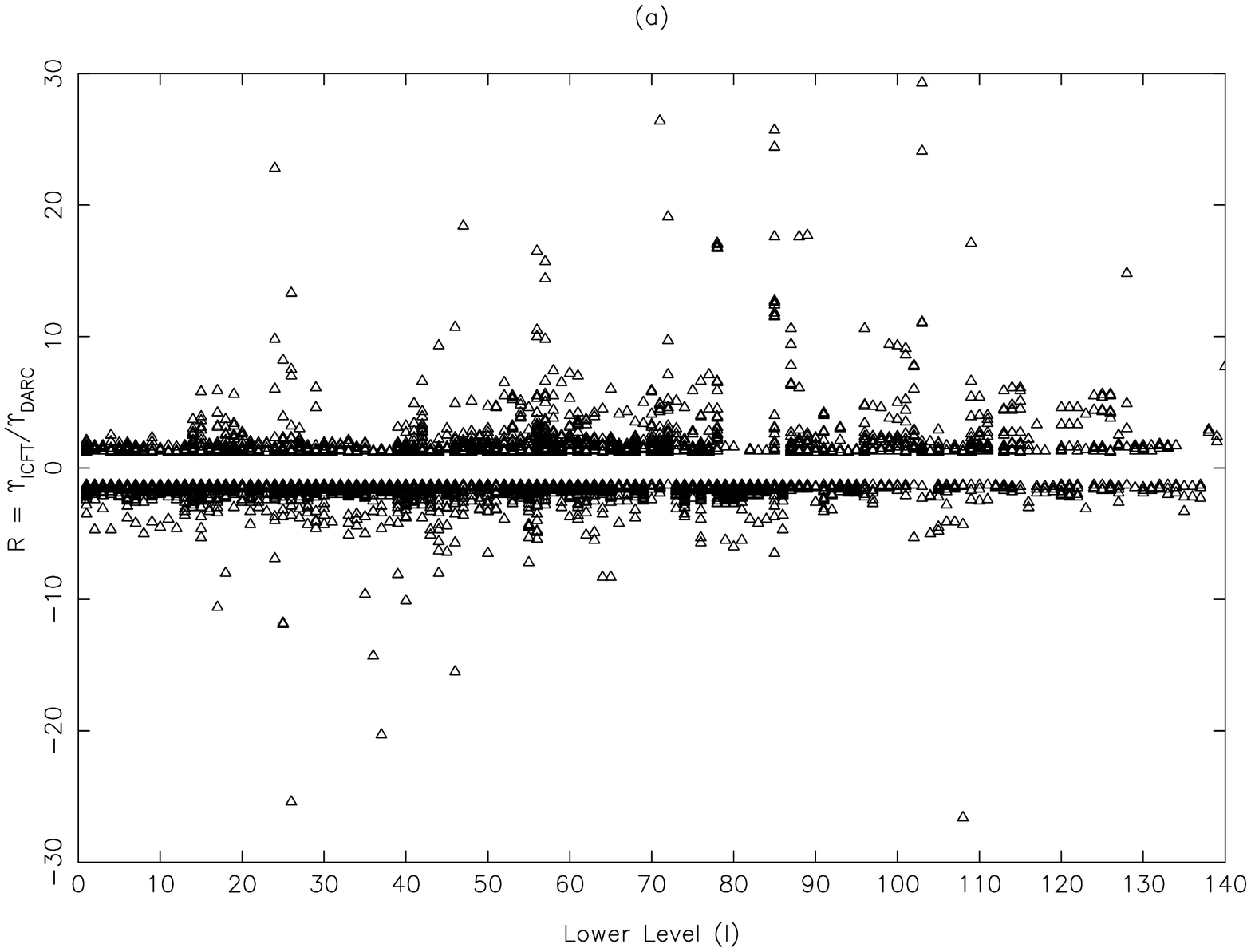}
 \vspace{-1.5cm}
% \caption{Comparison of DARC and ICFT values of $\Upsilon$ for transitions of Si~III at  T$_e$ = 1.8$\times$10$^6$~K. Negative R values plot  $\Upsilon_{\rm DARC}$/$\Upsilon_{\rm ICFT}$ and indicate that $\Upsilon_{\rm DARC}$ $>$ $\Upsilon_{\rm ICFT}$. Only those transitions are shown which differ by over 20\%. (a) Transitions {\em from} lower and (b) {\em to} upper levels.}
 \end{figure*}
 
\begin{figure*}
% \vspace{250pt}
\includegraphics[angle=-90,width=0.9\textwidth]{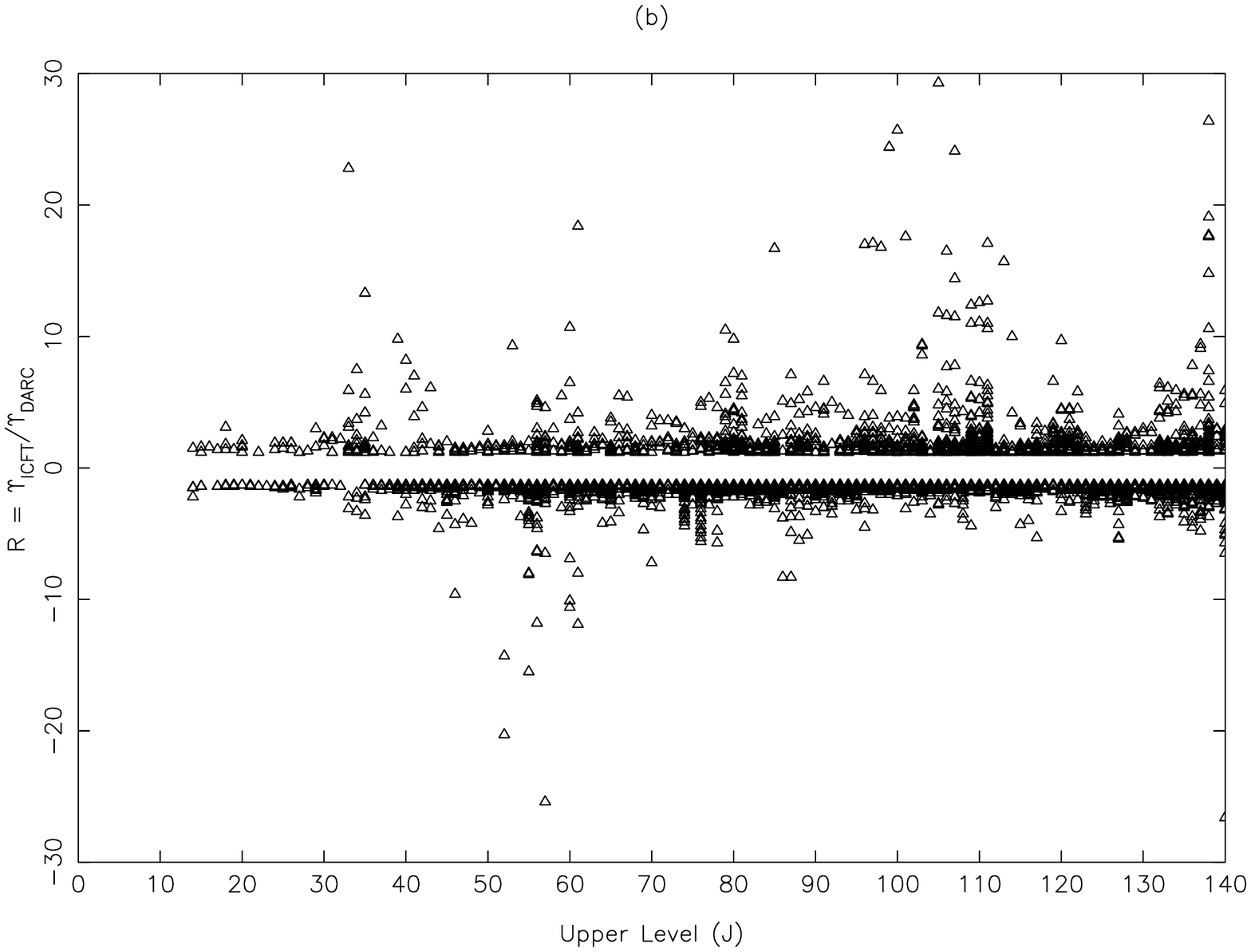}
 \vspace{-1.5cm}
 \caption{Comparison of DARC \cite{si3} and ICFT \cite{nrmglike}  values of $\Upsilon$ for transitions of Si~III at  T$_e$ = 1.8$\times$10$^6$~K. Negative R values plot  $\Upsilon_{\rm DARC}$/$\Upsilon_{\rm ICFT}$ and indicate that $\Upsilon_{\rm DARC}$ $>$ $\Upsilon_{\rm ICFT}$. Only those transitions are shown which differ by over 20\%. (a) Transitions {\em from} lower and (b) {\em to} upper levels.}
 \end{figure*}

Most of the discrepancies discussed above are between the DARC and ICFT calculations for $\Upsilon$. However, there are other examples also, apart from those discussed earlier in \cite{fst}. For the transitions of Mg~V, Hudson et al. \cite{hud} reported results among the lowest 37 levels of the 2s$^2$2p$^4$, 2s2p$^5$, 2p$^6$, 2s$^2$2p$^3$3s, and 2s$^2$2p$^3$3p configurations, for which they adopted the standard non-relativistic $R$-matrix code. Recently, Tayal and Sossah  \cite{sstmg5} extended these calculations, apart from making other improvements, to 86 levels of the  2s$^2$2p$^4$, 2s2p$^5$, 2p$^6$, and 2s$^2$2p$^3$3$\ell$ configurations. Based on comparisons for only a few transitions, they concluded  a good agreement for `most' transitions between their results of $\Upsilon$ and those of \cite{hud}. Unfortunately, this was not found  to be true as discussed in our work \cite{mgv}. In fact, the discrepancies between the two calculations are up to three orders of magnitude for over 80\% of the transitions, and in most cases the $\Upsilon$ of \cite{hud} are {\em higher} -- see Fig.~1 of \cite{mgv}. Furthermore, the discrepancies increase with increasing temperature, because \cite{hud} calculated  $\Omega$  only up to an energy of 28~Ryd, but determined $\Upsilon$  up to T$_e$ = 10$^7$~K, i.e. $\sim$63~Ryd, {\em without} any extrapolation of the $\Omega$ data.

\begin{figure*}
% \vspace{250pt}
\includegraphics[angle=-90,width=0.9\textwidth]{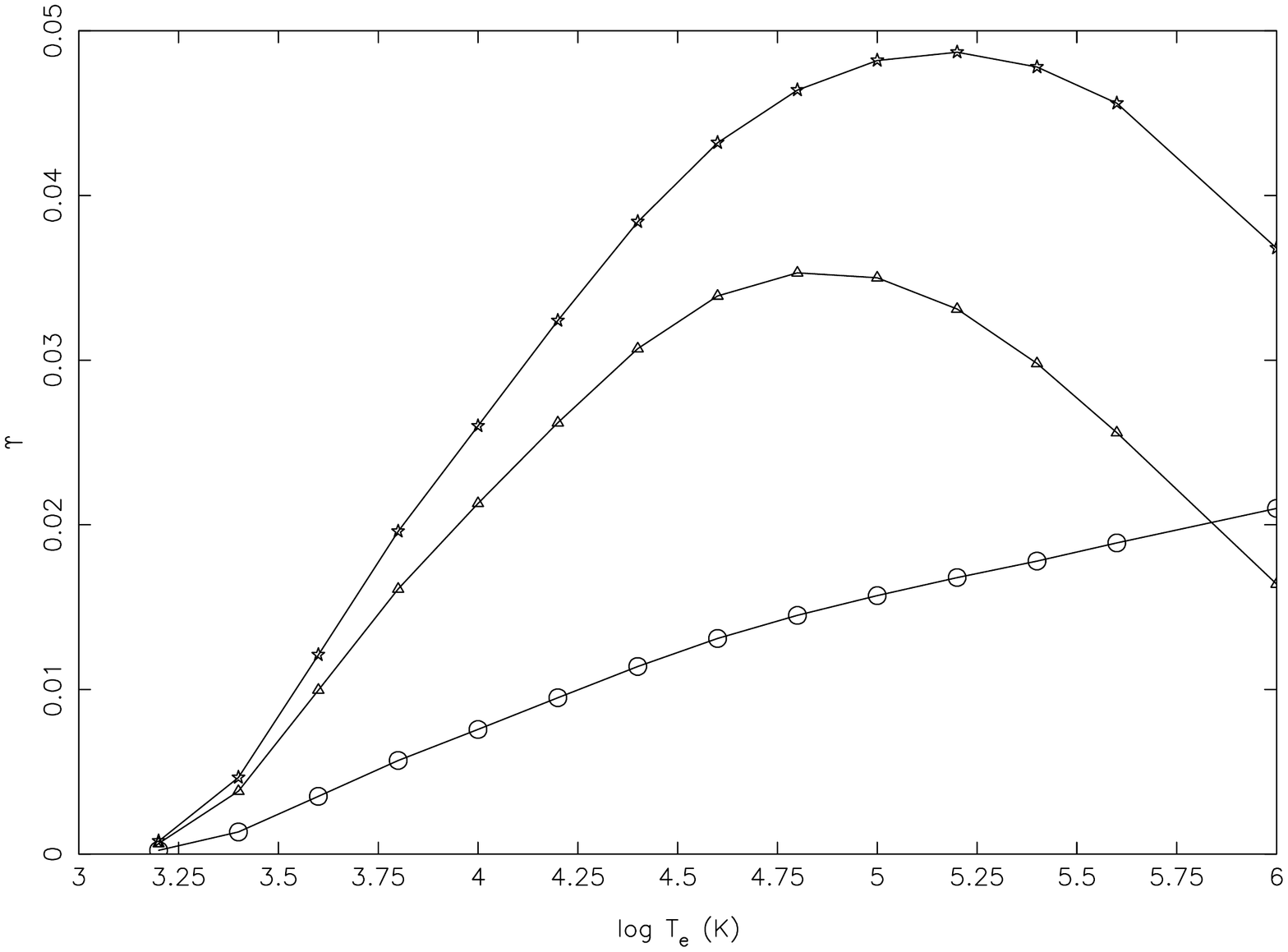}
 \vspace{-1.5cm}
 \caption{Effective collision strengths ($\Upsilon$) of \cite{sstmg5} for three {\em forbidden} transitions of Mg~V. Circles: 19--86 (2p$^3$3s~$^1$D$^o_2$ -- 2p$^3$($^2$P)3d~$^1$P$^o_1$), triangles: 24--86 (2p$^3$3s~$^3$P$^o_1$ -- 2p$^3$($^2$P)3d~$^1$P$^o_1$) and stars: 51--86 (2p$^3$($^4$S)3d~$^3$D$^o_3$ -- 2p$^3$($^2$P)3d~$^1$P$^o_1$) -- see \cite{mgv} for level indices. }
 \end{figure*}

Unfortunately, the $\Upsilon$ of \cite{sstmg5}  are also not correct for many transitions of Mg~V, and at almost all temperatures because of a variety of reasons, discussed and explained in \cite{mgv}. Particularly, at higher temperatures their $\Upsilon$ values increase with increasing temperature, irrespective of the (type of) transitions: allowed, inter-combination or forbidden. As an example, in Fig.~7 we show the variation of their $\Upsilon$ with T$_e$  for three {\em forbidden} transitions of Mg~V, namely 19--86 (2p$^3$3s~$^1$D$^o_2$ -- 2p$^3$($^2$P)3d~$^1$P$^o_1$),  24--86 (2p$^3$3s~$^3$P$^o_1$ -- 2p$^3$($^2$P)3d~$^1$P$^o_1$) and  51--86 (2p$^3$($^4$S)3d~$^3$D$^o_3$ -- 2p$^3$($^2$P)3d~$^1$P$^o_1$) -- see \cite{mgv} for level indices. For these (and many more) transitions their $\Upsilon$ values increase over a large range of T$_e$ whereas they should decrease as shown in table~3 of \cite{mgv}. Since level 86 (2p$^3$($^2$P)3d~$^1$P$^o_1$) is the highest threshold in their (or our) calculations, there should be no resonances related to this level, but the increasing behaviour of their $\Upsilon$ indicates the presence of pseudo resonances. 
 
This point has recently been contested by Wang et al. \cite{kbmgv}. Our speculation was based on the behaviour of $\Upsilon$ (as discussed in detail in \cite{mgv}), because the corresponding data for $\Omega$ were neither published nor were made available to us even after requests to the first author. However, Wang et al. \cite{kbmgv} had access to the $\Omega$ data of \cite{sstmg5}, and therefore they could diagnose to the problem of using a (very) coarse energy mesh close to the highest threshold (86)  of Mg~V, which affected the results shown in table~3 of \cite{mgv}. As a result,  not only transitions related to this threshold were affected but also others -- see for example, 19--78/81/82/84/85. Moreover, Wang et al. \cite{kbmgv} have speculated that both Hudson et al. \cite{hud} and our results \cite{mgv} of $\Omega$ (and subsequently $\Upsilon$) have pseudo-resonances, which is {\em not} the case. The reason for the higher values of $\Upsilon$ reported by \cite{hud} has been fully explained in our work \cite{mgv} and is because of the limited energy range adopted by them. Similarly, we can {\em confirm} that there are {\em no}  pseudo-resonances in our $\Omega$ or $\Upsilon$ data. Therefore, as was advised in \cite{fst} it is preferable if authors can publish (at least) some data for $\Omega$ so that differences in $\Upsilon$ results can be diagnosed. Alternatively, they should either make it available on a website  or should provide the data on request.

\section{Conclusions and Suggestions}

Numerous examples have been shown here for the large discrepancies noted for almost all atomic parameters of interest, mainly energy levels, A-values and $\Upsilon$. In a majority of cases it is difficult to assess the accuracy of the published data without performing independent calculations, either with the same (or similar) or a different method. Unfortunately, it is not always possible because enough computational resources or expertise may not be available. More importantly, calculations for $\Upsilon$ are particularly very computationally intensive for which large resources and months (if not years) of (computational) time are required. Since more and more levels are now being included in the scattering calculations, it is almost impossible to reproduce and verify the reported data with another  similar work. Therefore, it is necessary for the producers of atomic data not to make the mistakes in the first place, and to make as many comparisons as possible. A limited comparison (as discussed for the transitions of Mg~V) may lead to wrong conclusions, and subsequently to incorrect results. 

It is generally believed (and is mostly true)  that if the wavefunctions (read energy levels) are more accurate then so are the subsequent parameters, such as $\Omega$ and $\Upsilon$. Similarly,  a larger model should lead to more accurate results for $\Upsilon$, because of the inclusion of additional resonances. However, neither of this is (strictly) true as shown in this work --  and also see a very detailed work by Feldman \cite{uf} on a series of ions. More importantly, in all scattering calculations some compromises are made, because the number of CSFs, which are often included in the calculations of energy levels and A-values, are drastically reduced for computational reason. Therefore, the accuracy achieved for these parameters is generally higher than for $\Omega$ and $\Upsilon$. Nevertheless, most of the compromises made in reducing the CSFs are reasonable and do not lead to large discrepancies (of orders of magnitude) in the determination of $\Omega$ and $\Upsilon$.

Most of the advices for avoiding large discrepancies in the calculations of atomic data have already been enumerated in our earlier paper \cite{fst} and hence are not repeated again, because practically these remain the same. However,  large discrepancies and over a long period of time still remain for a series of ions between the DARC and ICFT results of $\Upsilon$. This is in spite of the best efforts made by various  workers, including those who generate data from these codes. As was emphasized earlier \cite{fst}, almost all working codes are under continuous process of development and improvements, and the errors (as and when detected) are corrected. Similarly, there is always scope for improvement over any calculation, either by improving the accuracy of the wavefunctions and/or enlarging the size by inclusion of additional levels, and this has recently been demonstrated by Si et al. \cite{si1,si2} for He-like ions and by Benda and Houfek \cite{jak} for atomic hydrogen.

We hope the noted discrepancies between the DARC and ICFT results will also be soon resolved  for the benefit of the users. Irrespective of the code adopted for a calculation,  it is more important how it is implemented.  All the compromises (and some are inevitable)  made in performing a calculation should be reasonable and up to a limit. For example, up to what value of partial waves the calculations for $\Omega$ should be performed strongly depends on the energy, otherwise the results may be under/over estimated as discussed in \cite{mgv, nixi, w66a}.   Similarly, up to what energy the calculations for $\Omega$ should be performed strongly depends on the temperature up to which the values of $\Upsilon$ are required. If the results for $\Omega$ are extrapolated over an enormously large energy range (as often is the case with ICFT calculations) then the subsequent results for $\Upsilon$ may be inaccurate. Finally, very large calculations (involving hundreds of levels) for any ion should only be performed when absolutely necessary. This is because the chances of errors in such calculations are more and these cannot be easily reproduced for verification and/or assessment.

It is encouraging to note that some efforts have already been made to resolve discrepancies in $\Upsilon$ data. For example, very recently Fern{\'a}ndez-Menchero et al. \cite{kbniv} have performed yet another calculation for N~IV by adopting the BSR (B-spline R-matrix) method. They have also highlighted a few reasons for the (possible) inaccuracies in the ICFT calculations, particularly for the weak spin changing transitions. Another point which we have repeatedly emphasised is the limited range  of energy often adopted in the ICFT calculations but extrapolated to a very large extent in order to calculate $\Upsilon$ up to very high temperatures. This exercise leads to incorrect behaviour of $\Omega$ at higher energies, at least for some transitions, as shown in Figs. 4 and 5. Like our calculations with DARC, Fern{\'a}ndez-Menchero et al. \cite{kbniv} have also adopted a large energy range to avoid this possible source of inaccuracy. We hope if similarly large energy range is adopted in the ICFT calculations then the discrepancies in $\Upsilon$ results may not be as striking as noted and demonstrated in this paper.

\section*{Acknowledgment}  

We thank Dr. Luis Fern{\'a}ndez-Menchero for his kindness in providing $\Omega$ data for a few transitions and some clarifications about the ICFT calculations.  We also thank a Referee for his various suggestions.

\end{document}